\documentclass[12pt,rmp,showkeys,amsmath,amssymb]{revtex4}
\usepackage[dvips,dvipdf]{graphicx}
\usepackage{epsfig}
\usepackage{amsmath, amsthm, amssymb}
\usepackage{natbib}
\usepackage{bm}
\begin{document}
\title{Local Hidden Variables Underpinning of Entanglement and Teleportation}
\author{A. Kalev}
\email{amirk@techunix.technion.ac.il}
\author{A. Mann}
\email{ady@physics.technion.ac.il}
\author{M. Revzen}
\email{revzen@physics.technion.ac.il}
\affiliation{Department of Physics, Technion-Israel Institute of Technology, Haifa 32000, Israel.\vspace{0.7in}}
\begin{abstract}
Entangled states whose Wigner functions are non-negative may
be viewed as being accounted for by local hidden variables (LHV).
Recently, there were studies of Bell's inequality
violation (BIQV) for such states in conjunction with the well
known theorem of Bell that precludes BIQV for theories that have
LHV underpinning.
We extend these studies to teleportation which is also based on entanglement.
We investigate if, to what extent, and under what
conditions may teleportation  be accounted for via LHV theory.
Our study  allows us to expose the role of various quantum
requirements.
These are, \textit{e.g.}, the uncertainty relation among non-commuting operators, and the no-cloning theorem which forces the complete elimination of the teleported state at its initial port.\vspace{0.4in}
\end{abstract}
%
\maketitle
\newpage
\thesection{$1.\;\;$\textbf{INTRODUCTION}}
\par
Entangled states were placed at the center of counterintuitive
predictions of quantum mechanics with the appearance of the
celebrated paper by Einstein, Podolsky and Rosen (EPR) in 1935
\cite{EPR}.
In the nineteen-sixties Bell analyzed theories that
could be underpinned with local hidden variables (LHV) and showed
that these theories must abide by certain inequalities
\cite{Bell1,Bell2} (known  as Bell's inequalities).
The EPR state, in the version introduced by Bohm \cite{Bohm},
most clearly allows a violation of the inequalities.
Bell's analysis is commonly interpreted to mean that quantum mechanics is a genuine ``non-classical'' theory in the sense that it cannot be underpinned with LHV.
These studies employed states which are maximally entangled.
Gisin in 1991 \cite{Gis} showed that Bell's inequality
violation (BIQV) is possible for all pure states which possess some
entanglement; again, this study  used spin (or spin-like)
entangled states.

\par
Ironically, it was noted by Bell \cite{Bell3} that the non-negative Wigner function \cite{Wig} for the \textit{original} EPR state might be viewed as providing LHV underpinning for measurements corresponding to linear combination of position and momentum for that state.
Thus, it would seem to imply that the Wigner function for this (maximally entangled) state provides a local classical model of the correlations!
Bell's considerations stimulated a considerable amount of
research in the problem \cite{BW,Ch,Gour}.
This research showed that BIQV \textit{can} be achieved even for a
non-maximally entangled state, the two-mode squeezed state (TMSS), although its Wigner function is \textit{non}-negative.
(For infinite squeezing this state reduces to the EPR state,
\textit{i.e.}, reaches maximal entanglement.)

\par
An extended discussion of the problem   noted that  having  LHV
underpinning for the wave-function is not sufficient for LHV
interpretation of quantal predictions; in addition, the
observables must also be accounted for via such LHV \cite{Rev}.
Thus, to underpin expectation values with LHV, the Wigner function
for the observables must take on their eigenvalues as its possible
values. The cases considered in the literature, wherein BIQV with
TMSS was allowed, did not satisfy this requirement and hence did
not introduce a counter-example to Bell's considerations.
Following \cite{Rev} we shall designate as  ``non-dispersive'' an
observable whose Wigner function takes on the eigenvalues of the
observable as its possible values; ``dispersive'' will  refer to
observables which do not have this property. Thus, only when
having a non-negative Wigner function for the wave-function
\textit{and} non-dispersive observables one may  interpret the
theory as being underpinned with LHV; then, of course, no BIQV is
possible.

\par
Another purportedly purely  quantum phenomenon associated with
entangled states is the possibility of teleportation \cite{BB,Vid,BK}.
When the entanglement is among widely separated degrees
of freedom (usually referred to as a ``quantum resource''),
manipulations in one locale plus a classical
transmission of information (to the other locale) allows setting up the degrees of freedom in the second locale  to
emulate the quantum state that was coupled to the system in the first locale.
The realization of teleportation was  originally interpreted as predicated on quantal reasoning, \textit{i.e.}, precluding LHV underpinning \cite{Zeil,Fur,Bow,Zha}.
A teleportation protocol that yields fidelity
greater than $50\%$ implies the involvement of some entanglement,
thence, apparently,  requiring quantal reasoning \cite{BK,BF}.
To ensure security of quantum fingerprints, a higher fidelity of $66\%$ would be required \cite{GG1}.
However, as stated above,  the involvement of entangled states by
itself does not necessarily preclude LHV underpinning.

\par
Interestingly, continuous variable teleportation \cite{Vid,BK,Fur,Bow,Zha} utilizes, as a quantum resource, an entangled state which is represented  by a \textit{non}-negative Wigner function.
Moreover, it utilizes only \textit{non}-dispersive observables, that is, observables which do not violate Bell's inequalities.
In this paper we address  the problem: 
Does teleportation of a quantum state always serve as an indisputable evidence for quantumness?
If  we give an example where teleportation of a quantum state, as a whole, in a ``single-shot'', allows LHV underpinning we have shown that it does not.
The above considerations may suggest that despite the fact that quantum teleportation must involve states which possess some entanglement \cite{BK}, when these states are represented by non-negative Wigner function, teleportation could be underpinned by LHV theory with no need to invoke the ``non-classicality'' of quantum mechanics.
In this paper we study the implications for teleportation of having entangled states that allow LHV underpinning, in particular the TMSS.

\par
An immediate and obvious requirement for possible classical interpretation for teleportation is that the quantum state to be teleported be such that its Wigner function is non-negative.
It can be shown that a non-negative Wigner function of a pure state is necessarily a Gaussian function \cite{Hud,Sch}.
Hence, to allow possible LHV underpinning, we consider  the teleported \textit{and} the resource's state represented by Gaussian distributions.
Then, we use the rules of the classical probability theory to formulate a teleportation protocol.
This protocol is a generalization of the standard continuous variable teleportation protocol \cite{Vid,BK}. 
Clearly, not every Gaussian distribution is a Wigner function, \textit{e.g.}, a general Gaussian does not necessarily obey the uncertainty relations.
Those distributions which may be viewed as Wigner functions of some quantum state are termed `physically realizable distributions'.
Otherwise, these  are merely  mathematical distributions  that  can \textit{not} be considered as a  representation of some physical  state \cite{Mann,EW}.
For physically realizable Gaussian distributions, the generalized protocol becomes the standard quantum protocol and gives a LHV underpinning for quantum teleportation.
Hence, we may conclude that teleportation of a pure quantum state does not always  assure a ``non-classical'' effect. 
The possibility for teleportation (and its meaning) is also studied for Gaussian distributions which do not obey the uncertainty relations (and hence do not represent physically realizable states).
Below, we show that there are \textit{non}-realizable Gaussian distributions which yield an efficient teleportation protocol.
   
\par
We note that a classical interpretation for quantum states in phase space is possible only if these quantum states are represented by  \textit{mixed} classical states.
A ``pure'' state is a state with zero entropy while a ``mixed'' state is a state with non-zero entropy.
Here, classical states relate to the Shannon entropy \cite{SN}, while quantum states relate to the von Neumann entropy \cite{VN}.
For example, the pure classical state $W(q,p)=\delta(q-q_0)\delta(p-p_0)$ represents a point in phase space, thence, its Shannon entropy  is zero.
However, the pure quantum state $W(q,p)=\frac{1}{\pi}e^{-(q^2+p^2)}$ (whose von Neumann entropy is zero) is `smeared' over the entire phase space; hence, from a classical viewpoint, it is a  mixed state (that is, its Shannon entropy is larger than zero)  which represents a joint probability distribution in $q$ and $p$.
Since \textit{any} Gaussian Wigner function is smeared over the entire phase space (and thus occupies a non-zero area there \cite{Sch}), it may be considered as a \textit{mixed} classical state (a pure classical state, as was pointed out above, occupies a point in phase space).
Of particular interest in this regard is  the classical interpretation of a \textit{perfect} teleportation of  a quantum state which is represented by a \textit{mixed} classical  state.
A perfect teleportation means that performing the protocol \textit{only once}  yields an output state (at the receiving port) equal to the original input state.
The standard quantum teleportation protocol becomes perfect when the state of the resource is maximally entangled \cite{BB,Vid,BK}.
In this case, the probabilities of \textit{any} further measurements on the \textit{single} (quantum) system, located at the receiving port, are completely determined by the input state, whether it is pure or mixed.
In this sense a mixed (or a pure) quantum state is  teleported via a \textit{single} measurement.
Hence, when a LHV underpinning is possible, it is convenient to interpret a mixed classical state
as  giving the propensity of a \textit{single}   system to yield an outcome of a certain kind \cite{Pop}.
For example, the mixed classical state $W(q,p)=\frac{1}{\pi}e^{-(q^2+p^2)}$ represents, by this interpretation, the joint propensity of the ``position'' ($q$) and ``momentum'' ($p$) variables of a \textit{single} particle system (in one degree of freedom) to obtain specific values: it has a Gaussian propensity to obtain any position and momentum values.

\par
The paper is organized as follows.
In the next section (Section $2$) we recall a few properties of the Wigner function that will be used to underpin  teleportation with LHV.
In Section $3$ we describe the generalized teleportation protocol and analyze it for different cases.
In  Section $4$ we discuss our conclusions.
A tentative conclusion of our analysis is that teleportation, in the cases considered, may be formulated by the rules of classical probability theory, and therefore may be accounted for by LHV theory (wherein the phase space variables play the role of LHV).\vspace{0.2in}

\thesection{$2.\;\;$\textbf{THE WIGNER FUNCTION}}
\par
In order to underpin  teleportation with LHV, we first recall a few
properties of the Wigner function \cite{Woo}.
It was shown  that (for spinless particles) quantum mechanics  can be formulated solely on the Wigner function formulation, and this formulation is equivalent to the density operator formulation \cite{Bak}.
In order to represent a quantum state in phase space, we must define the notions of ``position'' and ``momentum''.
For simplicity we first consider a particle system with one degree of freedom.
We introduce a basis of its Hilbert space:
$B_q=\{|q\rangle:q\in\Re\}$, which we arbitrarily
interpret as position basis ($\Re$ being the field of real numbers).
Given the position basis $B_q$, we introduce the
\textit{conjugate} momentum basis $B_p=\{|p\rangle:p=\in\Re\}$,
by means of the Fourier transform ($\hbar=1$):
\begin{eqnarray}\label{q2p}
|p\rangle=\frac{1}{\sqrt{2\pi}}\int_{-\infty}^{\infty}dq
e^{ipq}|q\rangle\;.
\end{eqnarray}
Associated with these bases one can define two complementary
observables, $\hat{q}$ and $\hat{p}$, such that
\begin{eqnarray}\label{QP}
\hat{q}|q\rangle=q|q\rangle\;,\quad\hat{p}|p\rangle=p|p\rangle\;.
\end{eqnarray}
For this system, the phase space is  a two-dimensional vector
space over the field of real numbers, $\Re$. Its axes are the
$c$-number variables  associated with the complementary
observables $\hat{q}$ and $\hat{p}$ (quadratures).

\par
The Wigner function, $W_Q(q,p)$, for a quantal operator $\hat{Q}$  is 
\begin{eqnarray}\label{WigQ}
W_Q(q,p)=\int_{-\infty}^{\infty}dx e^{-ipx}\langle
q+x/2|\hat{Q}|q-x/2\rangle\;.
\end{eqnarray}
For convenience, the Wigner function for the density operator $\hat{\rho}$ is defined with an extra factor $\frac{1}{2\pi}$ for each degree of freedom, \textit{i.e.}, for one degree of freedom
\begin{eqnarray}\label{WigRho}
W_\rho(q,p)=\frac{1}{2\pi}\int_{-\infty}^{\infty}dx
e^{-ipx}\langle q+x/2|\hat{\rho}|q-x/2\rangle\;.
\end{eqnarray}

\par
The Wigner function has a number of interesting properties, many
of which are discussed in Ref.\cite{Sch,Hill}. In this section, we
will mention just three special properties that will be important
for our purpose.
\par \textit{Property} (W1). For Hermitian operator $\hat{Q}$: $\forall(q,p)\in\Re^2$, $W_Q(q,p)\in\Re$.
\par \textit{Property} (W2). $\int dqdpW_{\rho}(q,p)=1$.
\par \textit{Property} (W3). Let $\hat{Q}$ and $\hat{Q}'$ be quantal
operators which act on the Hilbert space of the system, and let
$W$ and $W'$ be the corresponding Wigner functions. Then
$\frac{1}{2\pi}\int dqdp W_Q(q,p)
W_{Q'}(q,p)=Tr[\hat{Q}\hat{Q}']$. \\ Notice that if
$\hat{Q}=\hat{\rho}$, then the quantal expectation value of $\hat{Q}'$ is simply given by $\int dqdp W_\rho(q,p) W_{Q'}(q,p)$ [\textit{cf}. Eq(\ref{WigRho})].

\par
Properties (W1)-(W3) are crucial for the  LHV underpinning of entanglement.
Also, it is worth mentioning here two results which follow
immediately from Eq. (\ref{WigRho}):
$W_\rho(p)=\int W_\rho(q,p)dq$ is the probability density
for momentum, and $W_\rho(q)=\int W_\rho(q,p)dp$ is the probability
density for position.

\par
Although its marginals  are probability densities, in general the Wigner function does not have the meaning of a probability density.
It can take on negative values.
Nevertheless, a \textit{non}-negative Wigner function of a pure quantum  state is necessarily a (normalized) Gaussian \cite{Hud,Sch}, thus may be accounted for by the phase space coordinates as its LHV underpinning.

\par
We illustrate the above considerations using the TMSS defined as \cite{BW}
\begin{eqnarray}\label{TMSS}
&&|TMSS\rangle_{1,2}=\\\nonumber&&=\left(\frac{2}{\pi}\right)^{1/2}\int
dq_1dq_2\exp{\left[-\frac{e^{2r}}{2}(q_1-q_2)^2-\frac{e^{-2r}}{2}(q_1+q_2)^2\right]}|q_1\rangle_1|q_2\rangle_2\\\nonumber&&
\underset{r\rightarrow\infty}{\rightarrow}\;\int
dq|q\rangle_1|q\rangle_2=|EPR\rangle_{1,2}\;.
\end{eqnarray}
The state $|q\rangle_{i}$ is the position basis of $H^{(i)}$, the Hilbert space of system $i$ ($i=1,2$).
In the limit of the squeezing parameter $r$ increasing without limit, the TMSS approaches the (normalized) maximally entangled EPR state \cite{EPR}.
Its Wigner function, $W_{TMSS}$, is given by \cite{BW}
\begin{eqnarray}\label{WTMSS}
W_{TMSS}&=&\left(\frac{2}{\pi}\right)^2e^{-e^{2r}[(q_2-q_1)^2+(p_1+p_2)^2]}e^{-e^{-2r}[(q_1+q_2)^2+(p_2-p_1)^2]}\\\nonumber
&&\underset{r\rightarrow\infty}{\rightarrow}\;\frac{1}{2\pi}\delta(q_2-q_1)\delta(p_2+p_1)=W_{EPR}\;.
\end{eqnarray}
Although the TMSS (and  its maximal limit the EPR state) is an \textit{entangled} state, its Wigner function is \textit{non}-negative for all $q$'s and $p$'s.

\par
This property might suggest that entanglement, in this case, may be accounted for in terms of LHV.
As stated above, a non-negative Wigner function for the wave-function is not sufficient for LHV interpretation of quantal predictions \cite{Rev}.
The observables must be non-dispersive in order to underpin
quantal predictions by LHV theory.
Let the Wigner function for the wave-function be non-negative (as is the case for the TMSS and its maximal limit the EPR state), then, the quantum expectation value of  a non-dispersive observable $\hat{A}$ whose eigenvalues are $A(\lambda)=W_A(q,p)$ is given by:
\begin{eqnarray}\label{LHVBIQ}
&&\langle\hat{A}\rangle_{Quantum}=\int dqdp
W_{A}(q,p)W_\rho(q,p)\\\nonumber&=&\int
A(\lambda)Pr(\lambda)d\lambda =\langle A\rangle_{Classical}\;.
\end{eqnarray}
Eq. (\ref{LHVBIQ}) means that the expectation value of a non-dispersive observable $\hat{A}$ in a state whose Wigner function  is non-negative may be viewed as given by a \textit{local, classical} theory.
Such observables, obviously, would not violate Bell's inequalities \cite{Rev}.

\par
Teleportation is a quantum (\textit{i.e}., ``non-classical'') phenomenon associated with entangled states.
However, non-dispersive observables and an entangled state whose Wigner function is non-negative  have been utilized for  teleporting  an arbitrary quantum state \cite{Vid,BK}.
We shall see below that, when also the state to be teleported has a non-negative Wigner function, teleportation may be accounted for by LHV theory.

\par
To expose the nature of teleportation,  it is fruitful to generalize the standard (quantum) teleportation protocol \cite{Vid,BK} by considering both the teleported and the resource's state as represented by general Gaussian distributions (whether or not they are physically realizable distributions).
As pointed above, since we are interested in distributions that can be viewed as providing LHV underpinning when a physical realization is feasible, it is sufficient to consider only Gaussian distributions.
The generalized protocol is then  investigated under different limits (\textit{e.g.}, when the resource is a non-realizable pure classical state versus maximally entangled state).\vspace{0.2in}

\thesection{$3.\;\;$\textbf{THE GENERALIZED TELEPORTATION PROTOCOL}}
\par
In analogy to the standard protocol \cite{BB,Vid,BK}, consider three subsystems labeled by  $j=1,2,3$.
The aim is to teleport the unknown state of system $1$, which is characterized by an arbitrary Gaussian distribution $W_{in}(\alpha_1)$ [the notation $\alpha_i=(q_i,p_i)$ is used].
For this, the state of systems $2$ and $3$ (\textit{i.e.}, the state of the resource) is ``prepared'' in a correlated Gaussian state denoted by $W_{2,3}$.
We define a $100\%$ efficient protocol to be  that for which the distribution function of the output system at the end of the protocol is given by $W_{out}=W_{in}$.

\par
The most general Gaussian phase space distribution which
characterizes the state of the resource is \cite{EW}:
\begin{eqnarray}\label{G23}
G_{2,3}(\bm{\eta})=\frac{1}{(2\pi)^2\sqrt{detV}} e^{-\frac{1}{2}
\bm{\eta}^{\dagger}V^{-1}\bm{\eta}}\;,
\end{eqnarray}
where $\bm{\eta}=\bm{\xi}-\langle\bm{\xi}\rangle$; $\bm{\xi}$
designates the real phase space vector $(q_2,p_2,q_3,p_3)$; and
$\langle\circ\rangle$ stands for an average, such that for any
function $D(\bm{\eta})$
\begin{eqnarray}\label{Mean}
\langle D\rangle=\int d\bm{\eta} D(\bm{\eta})G_{2,3}(\bm{\eta})
\; ,
\end{eqnarray}
where $d\bm{\eta}\equiv dq_2dp_2dq_3dp_3$.  
The correlation property of the Gaussian distribution is completely determined by the positive $4\times 4$ real symmetric matrix - the
co-variance matrix - $V$, defined by 
\begin{eqnarray}\label{Vij}
V_{ij}=  \frac{1}{2}\langle(\eta_i \eta_j + \eta_j \eta_i) \rangle
\;.
\end{eqnarray}
It was shown   that any Gaussian distribution can be transformed (via squeezing and local linear unitary transformations) into a
standard form with $\langle\bm{\xi}\rangle=0$ and its co-variance
matrix may be written as \cite{Dua,Sim,EW}:
\begin{eqnarray}\label{CoVar1}
V=\left(%
\begin{array}{cccc}
  a & 0 & c_1 & 0 \\
  0 & a & 0 & c_2 \\
  c_1 & 0 & b & 0 \\
  0 & c_2 & 0 & b
\end{array}%
\right)\equiv\left(%
\begin{array}{cccc}
  \Delta^2 q_1 & 0 & \langle q_1q_2\rangle & 0 \\
  0 & \Delta^2 p_1 & 0 & \langle p_1p_2\rangle \\
  \langle q_1q_2\rangle & 0 & \Delta^2 q_2 & 0 \\
  0 & \langle p_1p_2\rangle & 0 & \Delta^2 p_2
\end{array}%
\right)\;.
\end{eqnarray}
Here the variance $\Delta^2$ of a phase space variable, $x$, is defined by 
\begin{eqnarray}\label{Var}
\Delta^2x=\langle x^2\rangle-\langle x\rangle^2 \; .
\end{eqnarray}

\par
Before moving on we note that the Gaussian distributions
(\ref{G23}) are not physically realizable for \textit{all} values
of the parameters $a,b,c_1,$ and $c_2$ of Eq. (\ref{CoVar1}).
The Gaussian distribution (\ref{G23}) may be physically realizable only in specific regions of the parameter space of $a,b,c_1,$ and $c_2$.
These regions are determined by the position-momentum uncertainty
relations \cite{Mann,EW}:  
\begin{eqnarray}\label{PMUncer}
&&\Delta^2q_2\Delta^2p_2=a^2\geq
1/4,\quad\Delta^2q_3\Delta^2p_3=b^2\geq 1/4\;,\\\nonumber
&&\Delta^2(q_3+q_2)\Delta^2(p_3+p_2)=(a+b+2c_1)(a+b+2c_2)\geq 1\;,\\\nonumber
&&\Delta^2(q_3-q_2)\Delta^2(p_3-p_2)=(a+b-2c_1)(a+b-2c_2)\geq 1\;.
\end{eqnarray}
The complementary regions:
\begin{eqnarray}\label{PMUncerV}
&&0\leq\Delta^2q_2\Delta^2p_2=a^2<
1/4,\quad 0\leq\Delta^2q_3\Delta^2p_3=b^2< 1/4\;,\\\nonumber
&&0\leq\Delta^2(q_3+q_2)\Delta^2(p_3+p_2)=(a+b+2c_1)(a+b+2c_2)< 1\;,\\\nonumber
&& 0\leq\Delta^2(q_3-q_2)\Delta^2(p_3-p_2)=(a+b-2c_1)(a+b-2c_2)< 1\;,
\end{eqnarray}
violate the uncertainty relations; therefore, in these regions the Gaussian distribution (\ref{G23}) is necessarily \textit{not} physically realizable,  \textit{i.e.}, it is not a Wigner function.

\par
The generalized teleportation protocol begins as follows:
The initial state of the system  in terms of its  phase space
distribution functions is:
\begin{eqnarray}\label{D123}
W_{1,2,3}&=&W_{in}(\alpha_1)W_{2,3}(\alpha_2,\alpha_3)\;.
\end{eqnarray}
$W_{2,3}$ is a standard form Gaussian, namely,
\begin{eqnarray}\label{D23}
W_{2,3}(\bm{\eta})=\frac{1}{(2\pi)^2\sqrt{detV}} e^{-\frac{1}{2}
\bm{\eta}^{\dagger}V^{-1}\bm{\eta}}\;,
\end{eqnarray}
where \begin{eqnarray}\label{CoVar2}
V^{-1}=\left(%
\begin{array}{cccc}
  b/(ab-c^2_1) & 0 & -c_1/(ab-c^2_1) & 0 \\
  0 & b/(ab-c^2_2) & 0 & -c_2/(ab-c^2_2) \\
  -c_1/(ab-c^2_1) & 0 & a/(ab-c^2_1) & 0 \\
  0 & -c_2/(ab-c^2_2) & 0 &  a/(ab-c^2_2)
\end{array}%
\right)\;.
\end{eqnarray} is the inverse matrix of $V$  and $detV=(ab-c^2_1)(ab-c^2_2)$.
For physically realizable  $W_{in}$ and $W_{2,3}$, Eq. (\ref{D123}) gives the phase space description for the initial quantum state.
However,  for  (general) Gaussian input and resource states, Eq. (\ref{D123}) has also a natural classical interpretation:
$W_{1,2,3}$ represents the  probability distribution of the composite system, and it is equal to a product of two probability distributions ($W_{in}$ and $W_{2,3}$) of  statistically independent subsystems.

\par
After preparing the initial state the  protocol proceeds as follows:
First, a measurement of the variables $q=q_2-q_1$ and $p=p_2+p_1$ is performed.
This measurement involves measurement  of classical currents \cite{BK}.
The probability density for getting a result $\beta=(q_\beta,p_\beta)$ is
\begin{eqnarray}\label{DBeta}
P(\beta)=\int
d^2\bm{\alpha}W_{in}(\alpha_1)W_{2,3}(\alpha_2,\alpha_3)\delta(q_2-q_1-q_\beta)\delta(p_2+p_1-p_\beta)\;,
\end{eqnarray} where $d^2\bm{\alpha}=\prod_{i=1}^{3}dq_idp_i$.
The classical expression for the probability, given in Eq. (\ref{DBeta}), becomes the quantal expression when the involved distributions are the Wigner distributions.
By definition, after the measurement, the (normalized) state of the third  subsystem is described by:
\begin{eqnarray}\label{D'3}
W'(\alpha_3|\beta)=\frac{1}{P(\beta)}\int
d^2\alpha_1d^2\alpha_2W_{in}(\alpha_1)W_{2,3}(\alpha_2,\alpha_3)\delta(q_2-q_1-q_\beta)\delta(p_2+p_1-p_\beta)\;.
\end{eqnarray}
Note that when the initial state of the system (\textit{i.e.}, $W_{1,2,3}$) is physically realizable, $W'(\alpha_3|\beta)$ is the quantal phase space description of the state of the third  subsystem (given a measurement outcome $\beta$).
For a general initial state, $W'(\alpha_3|\beta)$ has a classical interpretation:
It is  the  probability for the third subsystem  to be in phase space point $\alpha_3$ conditioned  by the measurement result  $\beta$.

\par
The final step of the protocol is to translate the third subsystem in phase space by $(-q_{\beta},p_{\beta})$ \cite{Vid,BK}.
Namely, 
\begin{eqnarray}\label{OutQP}
q_3\rightarrow q=q_3-q_{\beta}\;,\\\nonumber p_3\rightarrow
p=p_3+p_{\beta}\;.
\end{eqnarray}
In terms of the output variables, $\alpha=(q,p)$, the conditional probability distribution $W'$ is written as:
\begin{eqnarray}\label{D3beta}
W'(\alpha_3|\beta)&=&W'(q+q_{\beta},p-p_{\beta}|\beta)\equiv W_{out}(\alpha|\beta)\\\nonumber
&=&\frac{1}{P(\beta)}\int
d^2\alpha_1d^2\alpha_2W_{in}(\alpha_1)W_{2,3}(\alpha_2,q+q_{\beta},p-p_{\beta})\delta(q_2-q_1-q_\beta)\delta(p_2+p_1-p_\beta)\;.
\end{eqnarray}
An explicit expression for the conditional probability,
$W_{out}(\alpha|\beta)$, is obtained by using Eqs.(\ref{D23},\ref{CoVar2}) and performing an integration over
$\alpha_2$:
\begin{eqnarray}\label{D3beta2}
W_{out}(\alpha|\beta)&=&\frac{1}{P(\beta)}\frac{1}{2\pi\sqrt{detV}}\times\\\nonumber
&\times&\int
d^2\alpha_1W_{in}(\alpha_1)e^{-\frac{(q-q_1)^2}{2(a+b-2c_1)}
-\frac{(p-p_1)^2}{2(a+b+2c_2)}-\frac{(q+q_1+2q_\beta)^2}{2(a+b+2c_1)}-\frac{(p+p_1-2p_\beta)^2}{2(a+b-2c_2)}}\;.
\end{eqnarray}
The phase space distribution  function produced at the output of the teleportation device is given by averaging the conditional distribution $W_{out}(\alpha|\beta)$ over all possible measurement outcomes $\beta$:
\begin{eqnarray}\label{D3out}
W_{out}(\alpha)&=&\int d^2\beta
P(\beta)W_{out}(\alpha|\beta)\\\nonumber
&=&\frac{1}{2\pi\sqrt{(a+b-2c_1)(a+b+2c_2)}}\int
d^2\alpha_1W_{in}(\alpha_1)e^{-\frac{(q-q_1)^2}{2(a+b-2c_1)}
-\frac{(p-p_1)^2}{2(a+b+2c_2)}}\;.
\end{eqnarray}
This completes the protocol  for  teleportation of a phase space distribution.
Note  that Eq. (\ref{D3out}), which has a classical probability interpretation, becomes the quantal expression when the involved distributions are Wigner distributions.
We conclude that the (generalized) teleportation protocol, formulated by classical theory, becomes realizable (\textit{i.e.}, quantal) assuming that the \textit{total} initial Gaussian distribution is realizable.
Hence, the standard quantum teleportation protocol may be formulated by LHV theory assuming that the \textit{total} initial distribution is given by a non-negative Wigner function.

\par
Before moving on to analyze the protocol, let us  note the
following. 
First, the standard  teleportation protocol, being quantum, must abide by quantum requirements, \textit{e.g.}, the uncertainty relation among conjugate variables, and the no-cloning theorem \cite{noClo1,noClo2,noCloMix} which forces the complete elimination of the teleported state at its initial port. 
The generalized protocol, described above,  does not generally abide by these requirements.
In fact, below we give an example for an efficient teleportation
protocol which  \textit{violates} the uncertainty relation among
conjugate variables (of course it is merely a mathematical
procedure and cannot be realized  physically). 
We note that the generalized protocol abides by the no-cloning theorem. 
After the measurement, the state of the system at the sending port is represented by
$\frac{1}{2\pi}\delta(q_2-q_1-q_\beta)\delta(p_2+p_1-p_\beta)$
(where $q_i$ and $p_i$ are the phase space variables of subsystem
$i$, and $q_\beta$ and $p_\beta$ are the results of the
measurement). The state of the input system, subsystem $1$, is
obtained by integrating over the phase space variables of
subsystem $2$:
\begin{eqnarray}\label{D'1}
W(\alpha_1|\beta)=\frac{1}{2\pi}\int dq_2dp_2\delta(q_2-q_1-q_\beta)\delta(p_2+p_1-p_\beta)=\frac{1}{2\pi}\;.
\end{eqnarray}
Hence, in the generalized protocol (whether or not it is physically realizable) the original input state is completely eliminated at its initial port.
This is not an ``accident''. 
Recently it was  shown that  a protocol for broadcasting an arbitrary continuous classical distribution while leaving the original distribution unperturbed cannot be formulated \cite{Daff}.
Hence, the generalized teleportation protocol must abide by the no-cloning theorem for \textit{all} regions  of the parameter space of $a,b,c_1,$ and $c_2$ of Eq. (\ref{CoVar1}), including  the non-physical regions of the parameter space. 

\par
Second, as mentioned above, the quantum teleportation protocol is
perfect (\textit{i.e.}, the conditional quantum state at the receiving port  is equal to the original input state, after performing  only one measurement),  when a maximally entangled state is used as a resource \cite{BB,Vid,BK}. 
In the generalized protocol, a perfect teleportation means the following: 
For a general Gaussian resource $W_{2,3}$, the conditional  probability state resulting after \textit{one} measurement, $W_{out}(\alpha|\beta)$, is $\beta$ dependent, see Eq. (\ref{D3beta}). 
From Eq. (\ref{D3beta}) [or equivalently, Eq. (\ref{D3beta2})], it is easy to see that $W_{out}(\alpha|\beta)$ may be expressed as
\begin{eqnarray}\label{D3beta3}
W_{out}(\alpha|\beta)=\frac{1}{P(\beta)}G(\alpha,\beta)\;,
\end{eqnarray}
where $G$ is some Gaussian function. 
For special cases $G(\alpha,\beta)=P(\beta)W_{in}(\alpha)$, we get
$W_{out}(\alpha|\beta)=W_{in}(\alpha)$. 
This means that, for these special cases,  performing the protocol \textit{only once} is enough for the output probability state $W_{out}$ to be equal to the input state  $W_{in}$. 
We shall see that this result aligns with the quantal result, when the generalized protocol becomes realizable and the state of the resource is the  maximally entangled EPR state. 
It appears that (formal) perfect teleportation is not a unique feature of a maximally entangled (quantum) resource. 
Below we give an example for a perfect teleportation protocol which does not obey quantum  laws (hence it is \textit{not} realizable in Nature).

\par
The efficiency of the  protocol is quantified by the fidelity \cite{BK}
\begin{eqnarray}\label{Fid} F=2\pi\int d^2\alpha
W_{in}(\alpha)W_{out}(\alpha)\;.
\end{eqnarray}
For our analysis it is useful  to write down the explicit
expression for the fidelity for a coherent Gaussian distribution
$W_{in}(q,p)=\frac{1}{\pi}e^{-q^2-p^2}$ :
\begin{eqnarray}\label{DFidCoh}
F=\frac{1}{\sqrt{(a+b-2c_1+1)(a+b+2c_2+1)}}\;.
\end{eqnarray}

\par
A straightforward result is that the protocol is maximally efficient (\textit{i.e.}, the fidelity obtains its maximal value $1$) whenever $(a+b-2c_1)$ and $(a+b+2c_2)$ are  equal to zero.
It is noteworthy that there are \textit{non}-realizable distributions which satisfy this condition.
For example, the protocol can be maximally efficient when the inequalities (\ref{PMUncer}) are maximally violated, \textit{i.e.}, when they take on the value zero.
In this case, the state of the resource shared between the transmitting and receiving ports is:
\begin{eqnarray}\label{Pure}
W_{2,3}=\delta(q_2)\delta(p_2)\delta(q_3)\delta(p_3)\;.
\end{eqnarray}
This represents a \textit{pure} classical state in which the co-variance matrix is the null matrix.
(This state is non-realizable: Quantum mechanics precludes  such states.)

\par
Let us discuss briefly what would have been implied by such a protocol had it been realizable:
Given that the  resource is represented by Eq. (\ref{Pure}), the input state $W_{in}$ is actually being measured at the transmitting port and then reconstructed at the receiving port.
A measurement of the input state simply means that the phase space variables $q_1$ and $p_1$ are  measured, and the distribution of the results is given by $W_{in}(q_1,p_1)$.
Since subsystem $2$ is in a pure classical state $\delta(q_2)\delta(p_2)$, $q_2$ and $p_2$ are deterministically known.
Hence, the measurement of the variables $q_2-q_1$ and $p_2+p_1$ at the transmitting port yields the value of $q_1$ and $p_1$ (say, $q^{(1)}_1$ and $p^{(1)}_1$, respectively) according to ``their'' distribution function $W_{in}$.
These results are then sent to the receiving port.
After an appropriate translation in phase space, the state of the
output system at the receiving port is $W_{out}(q,p)=\delta(q-q^{(1)}_1)\delta(p-p^{(1)}_1)$.
This is a pure state which is, generally, different from $W_{in}$.
Therefore, \textit{performing the protocol only once  is not sufficient for reconstructing the input state}.
One must perform the protocol  many times (the word `many' is used here in its statistical context).
At the $i$-th time, the state of the output system at the receiving port is $W^{(i)}_{out}(q,p)=\delta(q-q^{(i)}_1)\delta(p-p^{(i)}_1)$, where $q^{(i)}$
and $p^{(i)}$ is the $i$-th measurement result at the transmitting port.
Here, the \textit{ensemble} description of the output states 
is given by the state $W_{in}$.

\par
Next, we  consider another case in which the protocol is maximally efficient.
In this case the resource is  a mixed classical state:
\begin{eqnarray}\label{QPure}
W_{2,3}=\frac{1}{2\pi}\delta(q_3-q_2)\delta(p_3+p_2)\;.
\end{eqnarray}
We note that  this state satisfies the quantum requirements [Eq. (\ref{PMUncer})].
It is the physically realizable pure EPR state \cite{We}.
For states that satisfy the quantum requirements, the generalized protocol becomes the standard (quantum) teleportation protocol \cite{Vid,BK}.
The standard protocol utilizes a pure quantum state as a resource to teleport a general (that is, a pure or mixed) state.
The EPR resource which is a \textit{pure} quantum state is represented in phase space by a \textit{mixed} classical state, \textit{i.e.}, by a classical distribution function
$W_{2,3}$.
We note that, given the EPR  resource, the resulting output state after performing the protocol only once is [see Eqs.(\ref{DBeta},\ref{D3beta})] $W_{out}(\alpha|\beta)=W_{in}(\alpha)$.
Hence, teleportation of a non-negative Wigner function $W_{in}$  \textit{as a whole via a single measurement} is formulated  by a classical theory, \textit{i.e.}, by a theory whose variables have definite values and for which it is possible to use the  rules of classical probability theory.
A convenient view for a probability distribution is in terms of the frequency a particular state occurs in an ensemble.
Here, perhaps, a more appealing view would be viewing the probability distribution as a \textit{single} system endowed with the propensity for the various outcomes of measurements.

\par
Let us discuss another case which may help us to understand the nature of teleportation.
Consider the mixed classical resource
\begin{eqnarray}\label{CEPR}
W_{2,3}=\frac{1}{2\pi}\delta(q_3-q_2)\delta(p_3-p_2)\;.
\end{eqnarray}
This state is clearly non-realizable.
Furthermore, this non-realizable state is related to the maximally entangled EPR state via $p_2\rightarrow-p_2$.
This is the Peres criterion for entangled states \cite{Per} in its version for continuous variables  bipartite Gaussian states \cite{Sim}.
Utilizing this state as a resource in the generalized protocol does not yield an efficient protocol [see Fig. ($1$)].
\begin{figure}[htbp]
\epsfxsize=.55\textwidth
\epsfysize=.35\textwidth  \centerline{\epsffile{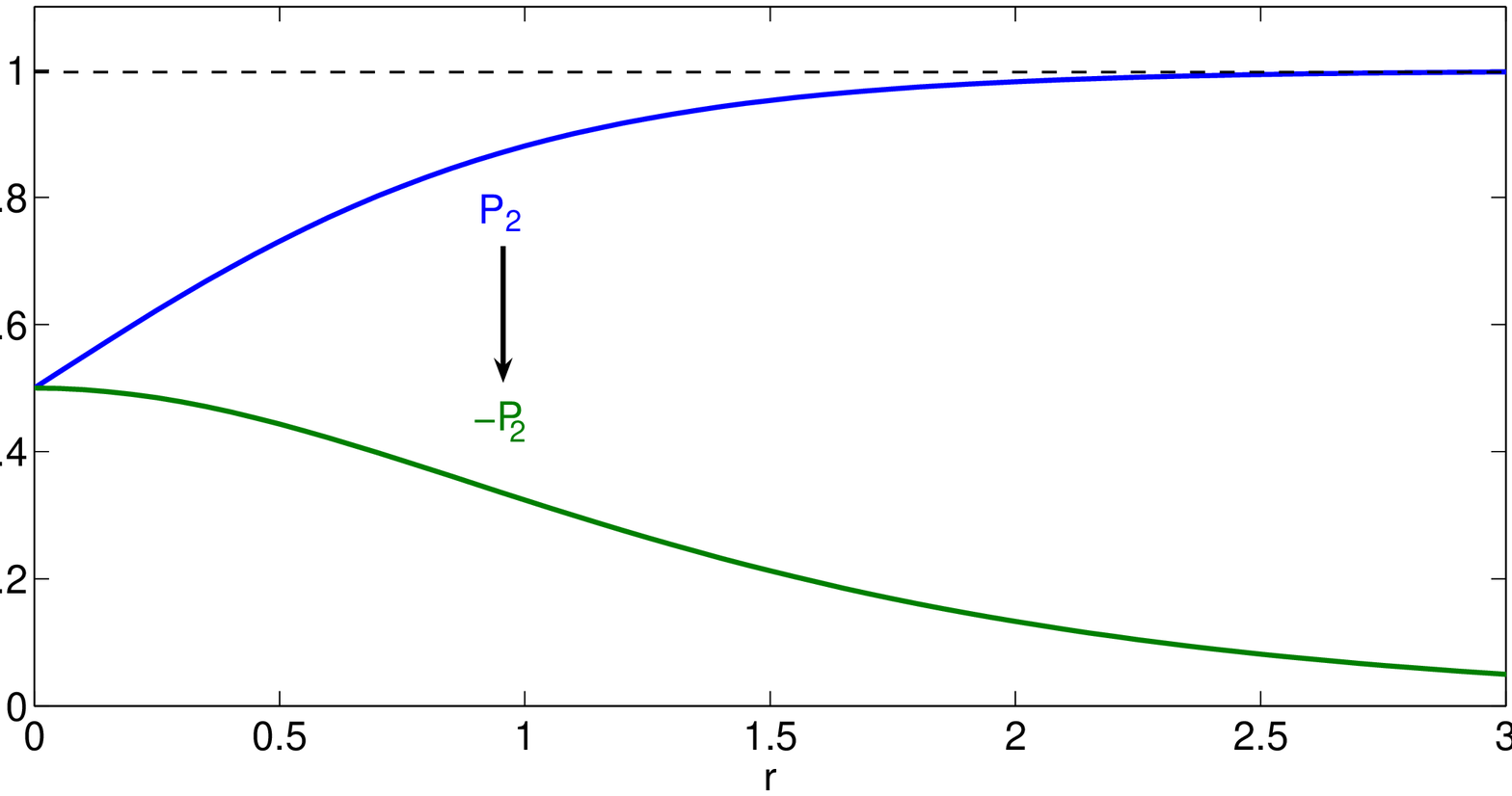}}
\vspace*{-0.2cm}
 \caption{\small{The fidelity of a coherent state teleportation. While the physically realizable protocol (upper line) succeeds to teleport the input state with fidelity $F=1/\left(1+e^{-2r}\right)$, the  non-realizable protocol (bottom line) fails in doing that and yields the fidelity $F=1/\sqrt{\left(1+e^{-2r}\right)\left(1+e^{+2r}\right)}\;.$} } \label{fig:Fid}
\end{figure}
However, there is an efficient \textit{classical} teleportation protocol which can (theoretically) use this state as a resource \cite{Br}.
The protocol is the same as the generalized protocol except for two points.
First, the variables measured at the transmitting port are $q_2-q_1$ and $p_2-p_1$
(instead of $q_2-q_1$ and $p_2+p_1$, as in the generalized protocol).
We note that no measuring technique is available for such measurement (this is a conjugate variables pair, and quantum mechanics prohibits a simultaneous measurement of conjugate variables).
Second, the phase space translations at the receiving port are
$q_3\rightarrow q=q_3-q_{\beta}$ and $p_3\rightarrow p=p_3-p_{\beta}$ [instead of the translations given in Eq. (\ref{OutQP})], where $q_{\beta}$ and
$p_{\beta}$ are the (would have been) measured values of $q_2-q_1$ and $p_2-p_1$, respectively.
It is easy to verify that, as in the previous case,  performing this protocol only once yields the resulting output state $W_{out}(\alpha|\beta)=W_{in}(\alpha)$.
Thus, had classical physics been realizable, teleportation of a (generally, mixed) classical state as a whole  in a single measurement would have been possible.
Nature obeys quantum physics rules, hence the only possible realization of teleportation is via quantum states which possess some entanglement.

\par
In recent studies \cite{Kon,Mor},  it was shown that (discrete) classical  probability distributions present  some interesting phenomena, one of which is closely related to teleportation (and usually referred to as classical ``one-time pad'').
Although the reasoning underlying these studies is not concerned with LHV, the conclusion is the same: Not all aspects of  teleportation are quantum.
The classical teleportation protocol that was presented above is a generalization of these studies to the case of continuous variables systems.\vspace{0.2in}  

\thesection{$4.\;\;$\textbf{DISCUSSION AND CONCLUSIONS}}
\par
The standard (physically realizable) teleportation protocol utilizes an entangled Gaussian state - the TMSS -  as a quantum resource \cite{BK}. 
The TMSS reduces to the  maximally entangled EPR state
in the limit of maximal squeezing [see Eq. (\ref{TMSS})]. 
Its Wigner function, Eq. (\ref{WTMSS}), is non-negative over the
whole phase space. 
We used this to view the  TMSS and the EPR state which are  \textit{pure} quantum states as  classical \textit{mixed} states.
A non-negative Wigner function of a state of a system is not sufficient  to allow a LHV account of measurements for other than  non-dispersive observables.
Measurements of dispersive observables on an entangled  Gaussian
state do not allow a local realistic description and thus can
violate Bell inequalities \cite{Rev}. 
We noted that the standard teleportation protocol with Gaussian input and resource uses only measurements of  observables which do not violate Bell inequalities. 
This means that teleportation of Gaussian states, although it must involve some entanglement, may be  accounted for in terms of a LHV theory with no need to invoke ``non-classical'' features of quantum mechanics. 
It should be clear that we do  not claim that teleportation of any quantum state can be underpinned by a classical theory. 
The main point we would like to establish is that there are  quantum states whose teleportation (within the standard protocol) has a ``classical'' description (our examples concern only states whose Wigner functions are non-negative). 
After teleportation is accomplished, these states could be used in various quantum tasks that may \textit{not} be described ``classically''. 
For example,  a ``classical'' description for teleportation of a Gaussian entangled state is valid, however, this state may be used  (at any time in the future) as a resource for  teleportation of a non-classical (e.g., number) state. 
It should be mentioned that a reasoning similar to \cite{Rev} was used in \cite{CW} where it was concluded that  classical interpretation for a Gaussian state teleportation is allowed.

\par
To show that the teleportation protocol with Gaussian input 
and resource could be formulated by a LHV theory, we have considered a  protocol which uses \textit{general} Gaussian distributions. 
Then, we followed the standard teleportation protocol \cite{Vid,BK}, and showed that  teleportation is obtained by using the  rules of classical probability theory.
Depending on the Gaussian's various parameters, we identified whether or not the protocol is physically realizable.

\par
The main conclusions of our study are:   
\begin{enumerate}
	\item Teleportation of a pure quantum state is not always an evidence for a ``non-classical'' phenomenon. The standard, quantum, protocol for teleporting  a non-negative Wigner function (utilizing a resource with a non-negative Wigner function) may be accounted for, in this case, by a LHV theory (wherein the phase space coordinates play the role of LHV). 
	\item When an EPR state, \textit{i.e.}, a maximally entangled pure state, is considered as a resource, the rules of classical probability theory are ``sufficient''  to formulate a $100\%$ efficient protocol that needs to be carried out \textit{only once} for teleporting an input state (that is, a non-negative Wigner function). For other resources, the protocol fails  to achieve maximal efficiency [Fig. ($1$)].
	\item A $100\%$ efficient protocol for  teleporting  classical states was formulated (theoretically): A maximally efficient protocol for teleporting an \textit{unknown} classical state via a single measurement was formulated when a mixed classical state is considered as a resource (this state is related to the maximally entangled EPR state by the Peres criterion \cite{Per} in its version for continuous variables bipartite Gaussian states \cite{Sim}). On the other hand, when a pure classical state is considered as a resource, the protocol must be carried out many times to achieve maximal efficiency. In this scenario, however, it ceases to function as a \textit{teleportation} protocol, since the input state is actually being measured at the sending port. 
	\item The generalized protocol  allowed us to view the role of  various quantum requirements in teleportation:
The uncertainty relation among conjugate phase space variables [Eq. (\ref{PMUncer})] and the no-cloning theorem \cite{noClo1,noClo2,noCloMix}.
We have seen that while the protocol (whether or not it could be physically realized) abides by the no-cloning theorem, it does not necessarily abide by the uncertainty relations.  This leads to a strict distinction between a realizable and a non-realizable teleportation protocol.
\end{enumerate}

\par
The representation of the realizable teleportation protocol in terms of classical probability distributions (\textit{i.e.}, mixed classical states) allows us to interpret the  classical probability theory in an ``untraditional'' way.
Traditionally, a mixed classical state  is interpreted as a state of some statistical ensemble.
Hence, the traditional interpretation suggests that the teleportation protocol may be accounted for via its mixed classical state representation by LHV of some statistical ensemble.
A realizable (\textit{i.e.}, quantum) state which is represented by a mixed classical state (as the TMSS and the EPR state) allows another interpretation:
A mixed classical state represents the propensity of the dynamical variables of a single system to obtain specific values (whether or not they can be measured simultaneously).
For example, the mixed classical state $W_{TMSS}$ is realized by a
\textit{pure} quantum state.
Thus for a \textit{single} physical system which is in a pure  TMSS,
$W_{TMSS}$ represents the propensity of the  dynamical
variables $q$ and $p$ to obtain specific values.
This interpretation suggests that the teleportation protocol may be
accounted for, via its mixed classical state representation,
by LHV of a \textit{single} physical system.\vspace{0.45in}

\textbf{ACKNOWLEDGMENTS}\\
We would like to thank  W. De Baere, N. Lindner, P. A. Mello, and in particular to S. L. Braunstein, for illuminating and helpful discussions.

\newpage\pagestyle{empty}
\begin{figure}[htbp]
\epsfxsize=.55\textwidth
\epsfysize=.35\textwidth  \centerline{\epsffile{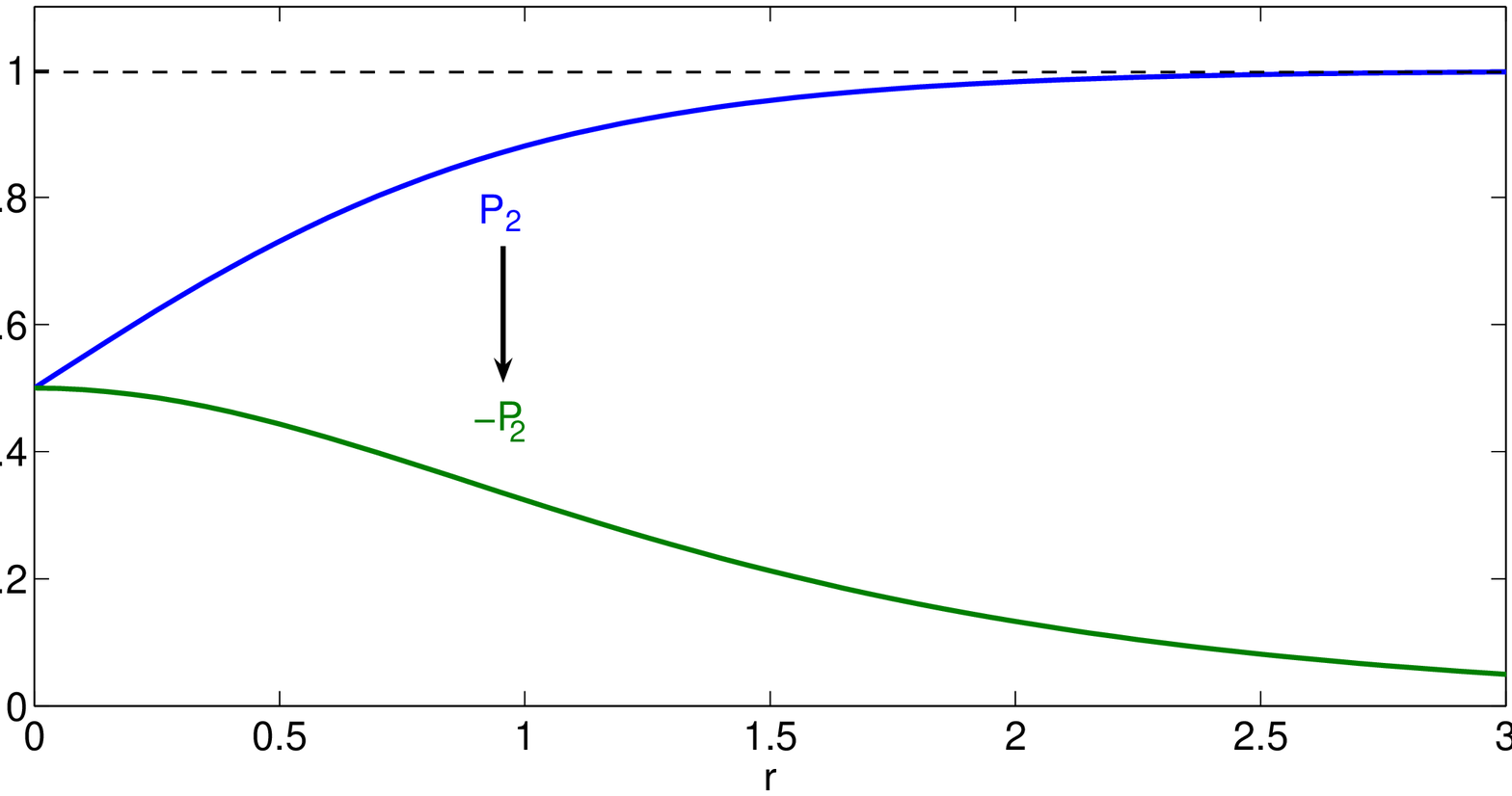}}
\end{figure}
\newpage\pagestyle{empty}
CAPTION:\\
The fidelity of a coherent state teleportation. While the physically realizable protocol (upper line) succeeds to teleport the input state with fidelity $F=1/\left(1+e^{-2r}\right)$, the  non-realizable protocol (bottom line) fails in doing that and yields the fidelity $F=1/\sqrt{\left(1+e^{-2r}\right)\left(1+e^{+2r}\right)}\;.$ 

\begin{thebibliography}{50}
\bibitem{We} We ignore the practical difficulty of realizing the pure EPR state.
Its realization requires an unbounded energy resource.
\end{thebibliography}
\end{document}